\documentclass[12pt,reqno]{amsart}
\usepackage{amsfonts}
\usepackage{amssymb}
\usepackage{amsmath}
\usepackage{cite}
\usepackage[dvips]{color}
\usepackage{pstricks}
\allowdisplaybreaks
\numberwithin{equation}{section}
\DeclareMathOperator{\sgn}{\rm sgn}
\DeclareMathOperator{\diag}{\rm diag}
\DeclareMathOperator*{\res}{\rm res}

\DeclareMathOperator{\CC}{\mathbb{C}}

\DeclareMathOperator{\Lo}{\mathcal{L}}
\DeclareMathOperator{\No}{\mathcal{N}}

\DeclareMathOperator{\Vo}{\mathcal{V}}

\DeclareMathOperator{\vhi}{\varphi}
\DeclareMathOperator{\la}{\lambda}
\DeclareMathOperator{\ka}{\varkappa}
\def\t#1{\widetilde{#1}}

\def\h#1{\widehat{#1}}

\newtheorem{theorem}{Theorem}[section]

\newtheorem{definition}{Definition}[section]
\newtheorem{remark}{Remark}[section]

\newtheorem{corollary}{Corollary}[section]
\newtheorem{condition}{Condition}[section]
\begin{document}

\title[Cauchy--Jost function and Darboux transformations]
{KPII: Cauchy--Jost function, Darboux transformations and totally nonnegative matrices}
\thanks{This work has been funded by the  Russian Academic Excellence Project '5-100'. Research has also been supported by 
the RFBR grants No.\ 14-01-00860 and No.\ 15-31-20484.}
\author{M.~Boiti, F.~Pempinelli and A.~K.~Pogrebkov}
\address{Steklov Mathematical Institute and National Research University Higher Scho\-ol of Economics, 
Moscow, Russian Federation, pogreb@mi.ras.ru;
EINSTEIN Consortium, Lecce, Italy}
\begin{abstract}
Direct definition of the Cauchy--Jost (known also as Cauchy--Baker--Ak\-hi\-e\-zer) function in the 
case of pure solitonic solution is given and properties of this function are discussed  in detail using the 
Kadomtsev--Petviashvili II equation as example. This enables formulation  of the Darboux transformations in terms of the 
Cauchy--Jost function and classification of these transformations. Action of Darboux transformations on Grassmanians---i.e., 
on the space of soliton parameters---is derived and relation of the Darboux transformations with property of total 
nonnegativity of elements of corresponding Grassmanians is discussed.
\end{abstract}
\maketitle


\section{Introduction and notation}
In this article we continue the investigation of the Cauchy--Jost function and demonstrate its exceptional role in 
the study of inverse problems. This function naturally appeared in the theory of (binary) B\"acklund transformations, 
see \cite{MS}. In \cite{GO} this  function was studied from an algebro-geometric point of view and it was called there the 
Cauchy--Baker--Akhiezer kernel. In \cite{BK1} and \cite{BPP2015} this function was studied by means of the 
$\bar\partial$-method for a different classes of the scattering data. It was proved that by means of this function all 
equations of the Kadomtsev--Petviashvili hierarchy can be presented in a compact  quasilinear form and Jost equations 
appear as  specific asymptotic values of the Cauchy--Jost function, that motivates our choice of its label.  This proves 
that this function is the most convenient object for investigation of the inverse problem. We take the  
Kadomtsev--Petviashvili II (KPII) equation, \cite{KP1970},
\begin{equation}
(u_{t}-6uu_{x_{1}}+u_{x_{1}x_{1}x_{1}})_{x_{1}}=-3u_{x_{2}x_{2}},\label{KPII}
\end{equation}
with Lax operator, \cite{D1974,ZS1974},
\begin{equation}
\Lo(x,\partial_{x})=-\partial_{x_{2}}+\partial_{x_{1}}^{2}-u(x),\label{heatop}
\end{equation}
as example. In this case the Cauchy--Jost function can be defined as a primitive of the product of the Jost and dual 
Jost solution:
\begin{equation}
F(x,\la,\mu)=\int\limits^{x_1}_{\pm\infty}dy_{1}\psi(y,\la)\vhi(y,\mu)\bigr|_{ y_2=x_2},\label{new1}
\end{equation} 
where $\vhi(x,\la)$ is the Jost solution of the heat equation $\Lo(x,\partial_{x})\vhi(x,\la)=0$ and $\psi(x,\la)$ is the 
Jost solution of its dual equation, $\la$ and $\mu$ are two complex parameters, $x=(x_1,x_2,x_3,\ldots)$ (up to some finite 
number) denotes space and time independent variables. The sign of the bottom limit of the integral in (\ref{new1}) must be 
chosen in order to guarantee its convergence. 

It is known, see \cite{GO}--\cite{BPP2015} for detail, that the Cauchy--Jost function has unity residual at $\mu=\la$ and 
decays when $\la$ or $\mu$ tend to infinity if the exponential factor $e^{(\ell(\la)-\ell(\mu))x}$, where
\begin{equation}
\ell(\la)x=\sum_{k\geq1}\la^{k}x_k,\label{new5} 
\end{equation} 
is removed. In particular in \cite{BK1} it is shown that in the case considered there this function is given in terms of the 
$\tau$-function as
\begin{equation}
F(x,\la,\mu)=\dfrac{e^{(\ell(\la)-\ell(\mu))x}}{\mu-\la}\dfrac{\tau(x-[\la^{-1}]+[\mu^{-1}])}{\tau(x)},\label{new2} 
\end{equation} 
where dependence on the infinite set of times $x=(x_1,x_2,x_3,\ldots)$ is assumed and square brackets denote the infinite 
Miwa vector
\begin{equation}
[\la]=\bigl(\la,\la^{2}/2,\la^{3}/3,\ldots\bigr).\label{new4} 
\end{equation} 

In \cite{BPP2015} the direct definition of the Cauchy--Jost function in the case of rapidly decaying potential was given by 
means of the $\bar\partial$-problem for it. The Jost solutions, potential and other objects are given then in terms of 
this function by proper limiting procedures. In particular, one can use an arbitrary finite number of variables $x_k$ in 
(\ref{new5}). Then evolution of the Cauchy--Jost function with respect to these times is given in a closed form and 
evolutions of the Jost solutions and potential follows consequently. Here we generalize the definition of the Cauchy--Jost 
function for the pure soliton potentials of the heat equation, investigate its properties, define Darboux transformations 
in its terms. This enables us to derive action of these transformations on points of Grassmanians that parametrize 
soliton potentials. 

The  article is organized as follows. In Sec.\ \ref{old} we give some notations and known results for the pure soliton 
potentials of the heat equation, see e.g.\ \cite{BPPPr2001a}--\cite{K} for detail. In particular, it is known  that soliton 
potentials are labeled by two numbers $N_{a}$ and $N_{b}$, so it is natural to speak about $(N_a,N_b)$-solitons.  
In Sec.\ \ref{CJ} we define the Cauchy--Jost function by means of its analyticity properties with respect to either one of 
the spectral parameters, $\la$ or $\mu$, and prove that these properties uniquely define function $F(x,\la,\mu)$ that in its 
turn defines the Jost solutions and potential of the heat equation in the pure solitonic case. In this sense our approach to 
the definition of the Cauchy--Jost function is close to that of \cite{generalKPsolitons} for the definition of the Jost 
(Baker--Akhiezer) solutions of the nonstationary Schr\"{o}dinger equation. In Sec.\ \ref{Darboux}we introduce discrete 
transformations of the Cauchy--Jost function given by quasilinear equations on it and show that they are equivalent to the 
original (bilinear) Darboux transformations of \cite{MS} and specify the latter ones. In accordance to two numbers 
labeling soliton solutions we single out three types of Darboux transformations: changing $N_a\to{N_a}+1$ and 
preserve $N_b$, vice verse, changing  $N_b\to{N_b}+1$ and preserve $N_a$ and those that does not change any of these numbers, 
but change other parameters of solitons. Correspondingly, $N_a$-transformations, $N_b$-transformations and 
0-transformations as well as transformations inverse to these ones. For any of these transformations we derive their 
action generated on the corresponding Grassmanian. Discussion of these results and their possible generalizations, as 
well as concluding remarks are given in Sec.\ \ref{discussion}. In Appendix \ref{app1} we present some relations of linear 
algebra used in our construction. 

\section{Jost solutions and potentials of the KPII equation in the pure solitonic case}\label{old}
We start with some notations used below. Let we have two numbers $N_{a}$ and $N_{b}$ obeying condition 
\begin{equation}
N_{a},N_{b}\geq 1. \label{nanb}
\end{equation}
Let 
\begin{equation}
\No=N_{a}+N_{b},\text{ so that } \No\geq 2. \label{Nnanb}
\end{equation}
We need $\No$ real parameters 
\begin{equation}
\ka_{1}<\ka_{2}<\ldots <\ka_{\No}, \label{kappas}
\end{equation}
and denote
\begin{align}
&\ell(\la)x=\la{x_1}+\la^{2}x_2+\la^{3}x_3+\ldots,\label{ellx}\\
&\ell_{n}x=\ka_{n}^{}x_1+\ka_{n}^{2}x_2+\ka_{n}^{3}x_3+\ldots,\quad n=1,\ldots,\No.\label{Kn} 
\end{align}
where $\ldots$ denote higher terms, if necessary, up to some finite order and where $\la\in\CC$. We use special diagonal real $\No\times{\No}$ matrices 
\begin{align}
&e^{\ell {x}}=\diag\{e^{\ell_{n}x}\}_{n=1}^{\No} \label{eK}\\
&r=\diag\{r_{n}\}_{n=1}^{\No},\qquad r_{n}=\prod_{\substack{n'=1 \\ n'\neq {n}}}^{\No}(\ka_{n}-\ka_{n'}),\label{r}\\
&\ka=\diag\{\ka_{1},\ldots,\ka_{\No}\}.\label{kdiag} 
\end{align}

In terms of parameters $\ka_{j}$ we introduce two ``incomplete Vandermonde matrices,'' i.e., $N_{b}\times\No$ and 
$\No\times{N_{a}}$ matrices
\begin{equation}
\Vo=\left(\begin{array}{lll}
1 &\ldots & 1\\ 
\ka_{1} &\ldots &\ka_{\No}\\ 
\vdots &  &\vdots\\ 
\ka_{1}^{N_{b}-1} &\ldots &\ka_{\No}^{N_{b}-1}
\end{array}\right),\qquad
\Vo^{\,\prime}=\left(\begin{array}{lll}
1 &\ldots &\ka_{1}^{N_{a}-1}\\ 
\vdots &  &\vdots\\ 
1 &\ldots &\ka_{\No}^{N_{a}-1}
\end{array}\right),  \label{V}
\end{equation}
that obey condition of ``orthogonality''
\begin{equation}
\Vo{r^{-1}}\Vo^{\,\prime}=0.\label{d12'} 
\end{equation} 
This relation follows from the trivial equality
\begin{equation}
\sum_{n=1}^{\No}\dfrac{\ka_{n}^{m-1}}{r_n}=\delta_{m\No},\quad m=1,\ldots,\No,\label{pol14} 
\end{equation} 
that coincides with (\ref{d12'}) for $m=1,\ldots,\No-1$. 

Finally, let $C$ be a $\No\times {N_{b}}$ real constant matrix. For maximal minors of the rectangular matrices introduced 
above we use notation of the kind
\begin{align}
&\Vo(\{n_{i}\})=\det\left(\begin{array}{lll}
1 &\ldots & 1\\ 
\ka_{n_{1}} &\ldots &\ka_{n_{N_{b}}}\\ 
\vdots &  &\vdots\\ 
\ka_{n_{1}}^{N_{b}-1} &\ldots 
&\ka_{n_{N_{b}}}^{N_{b}-1}\end{array}\right)=\prod_{1\leq{i}<j\leq{N_b}}(\ka_{n_j}-\ka_{n_i}),\label{Vo}\\
&C(\{n_{i}\})=\det\left(\begin{array}{ccc}
C_{n_{1},1} &\dots &C_{n_{1},N_{b}}\\ 
\vdots &  &\vdots\\ 
C_{n_{N_{b}},1} &\dots &C_{n_{N_{b}},N_{b}}\end{array}\right), \label{Do}
\end{align} 
where $\{n_1,\ldots,n_{N_b}\}\subset\{1,\ldots,\No\}$. In these terms we impose on matrix $C$ the following condition.
\begin{condition}\label{cC}
For every $n$, $1\leq{n}\leq\No$, there exists a subset $\{n_1,\ldots,n_{N_b-1}\}$ of $\{1,\ldots,\No\}$ such that
\begin{equation}
 C(n,n_1,\ldots,n_{N_b-1})\neq0.\label{condC}
\end{equation}
\end{condition}
Let $C'$ be a constant nonzero $N_{a}\times\No$ matrix  that is ``orthogonal'' to the matrix $C$ in the sense that in 
analogy to (\ref{d12'}) it obeys
\begin{equation}
C'rC=0, \label{d12}
\end{equation}
where zero in the r.h.s.\ is the $N_{a}\times {N_{b}}$ matrix. It is clear that thanks to (\ref{nanb}) matrix $C'$ exists 
and obeys condition above, as follows from consideration in Appendix~\ref{app1}. Also from relations in this appendix we get 
that matrices $C$ and $C'$ are complementary in the sense that for any 
$\No$-vector $w$ there exist such $N_b$-vector $w_{b}$ and $N_a$-vector $w_{a}$ that
\begin{equation}
w=Cw_{b}+r{C'}^{\text{T}}w_{a},\label{new6} 
\end{equation} 
where $\text{T}$ denotes transposition.

Below essential role is played by matrices $\Vo e^{\ell{x}}C$ and $C'e^{-\ell{x}}\Vo^{\,\prime}$. In 
particular, $\tau$-functions are given by determinants
\begin{equation}
\tau(x)=\det\bigl(\Vo e^{\ell{x}}C\bigr),\qquad\tau'(x)=
\det\left(C'e^{-\ell{x}}\Vo^{\,\prime}\right) . \label{tau}
\end{equation}
By means of the Binet--Cauchy formula for the determinant of a product of matrices
\begin{align}
&\tau(x)=\sum_{1\leq{n}_1<\ldots<n_{N_b}\leq\No}C(\{n_{i}\})\Vo(\{n_{i}\})\prod_{j=1}^{N_{b}}e^{
\ell_{n_{j}}x}_{}, \label{tauf1}\\
&\tau'(x)=\sum_{1\leq{n}_1<\ldots<n_{N_a}\leq\No}C'(\{n_{i}\})\Vo'(\{n_{i}\})\prod_{j=1}^{N_{a}}e^{
-\ell_{n_{j}}x}_{}.\label{tauf2}
\end{align}
The $(N_a,N_b)$-soliton solution is given by the standard equalities
\begin{equation}
 u(x)=-2\partial_{x_1}^{2}\log\tau(x)\equiv- 2\partial_{x_1}^{2}\log\tau'(x),\label{u}
\end{equation}
where both expressions coincide as these $\tau$-functions are proportional (see Appendix \ref{app1}):
\begin{equation}
\tau(x)=\text{const}\cdot\biggl(\prod_{n=1}^{\No}e^{\ell_nx}\biggr)\tau'(x).\label{tautau} 
\end{equation} 
Relation (\ref{tauf1}) also clarifies meaning of the Condition~\ref{cC}: if there exists such $n_0$ that 
for any $n_1,\ldots,n_{N_b-1}$ all maximal minors $C(n_0,n_1,\ldots,n_{N_b-1})=0$ then $\tau(x)$ is independent of 
$\ka_{n_0}$. Thus in this case we have either $(N_a-1,N_b)$-, or $(N_a,N_b-1)$-soliton  solution given by $\No-1$ parameters 
$\ka_{j}$. Thanks to (\ref{tauf2}) the same is valid for the matrix $C'$. 

Relation (\ref{u}) demonstrates well known fact that potential $u(x)$ in the pure soliton case is invariant under 
transformations
\begin{equation}
 C\to Cc,\qquad C'\to c'C',\label{gr}
\end{equation}
where $c$ and $c'$ are arbitrary nondegenerate matrices of the size $N_b\times{N_b}$ and $N_a\times{N_a}$ 
correspondingly. Thus soliton solutions of the KPII equation are parametrized by points of real Grassmanian $Gr_{N_b,\No}$, 
or $Gr_{N_a,\No}$.

In what follows we also need the $\No$-th order polynomial
\begin{equation}
R(\la)=\prod_{n=1}^{\No}(\la-\ka_{n}),\label{Rla} 
\end{equation} 
so that by (\ref{r}):
\begin{equation}
r_n^{}=\dfrac{dR(\la)}{d\la}\biggr|_{\la=\ka_{n}}
\text{ and }\sum_{n=1}^{\No}\dfrac{\ka_n^j}{(\la-\ka_n)r_n}=\dfrac{\la^j}{R(\la)},\quad j=0,\ldots,\No-1.\label{Rr} 
\end{equation} 
Denoting symmetric polynomials of $-\ka_{1},\ldots,-\ka_{\No}$ as $s_{n}=s_{n}(-\ka_{1},\ldots,\newline-\ka_{\No})$, where
\begin{equation}
s_{m}(x_1,\ldots,x_{N})=\sum_{1\leq n_1<\cdots<n_m\leq N}x_{n_1}\ldots x_{n_m},\qquad s_{0}=1,\label{a-6}
\end{equation}
we can also write
\begin{equation}
R(\la)=\sum_{m=0}^{\No}\la^{\No-m}s_{m}.\label{Rla4} 
\end{equation} 
Taking into account that any $\ka_{n}$ is a root of $R(\la)$ we get
\begin{align}
\dfrac{R(\la)}{\la-\ka_{n}}=\sum_{j=0}^{\No-1}R_{\No-1-j}(\la)\ka^{j}_{n},\label{Rla5} 
\end{align}
where
\begin{equation}
R_{j}(\la)=\sum_{i=0}^{j}\la^{j-i}s_{i}\equiv\left(\dfrac{R(\la)}{\la^{\No-j}}\right)_{+}.\label{Rla6} 
\end{equation}
It is easy to see, that
\begin{equation}
R(\la)-\la^{\No-j}R_{j}(\la)=\sum_{i=j+1}^{\No}\la^{\No-i}s_{i},\label{new7} 
\end{equation} 
i.e., equals to a polynomial of order $\No-j-1$.

\section{Cauchy--Jost function}\label{CJ}
Let us consider the case where $k\geq3$ times are switched on, so we write $x=(x_1,\ldots,x_k)$ and, cf.\ (\ref{ellx}), $\ell(\la)x=\sum_{j=1}^{k}\la^{k}x_k$. Then using the above notation we set the following  definition. 
\begin{definition}\label{def1}
We say that function $F(x,\la,\mu)$, where $\la$ and $\mu$ are in $\CC$, is the Cauchy--Jost 
function of ($N_a,N_b$)-soliton solution of the KPII equation if function
\begin{equation}
f(x,\la,\mu)=e^{(\ell(\la)-\ell(\mu))x}F(x,\la,\mu)\label{new8} 
\end{equation} 
is such that product $(\la-\mu)f(x,\la,\mu)$ is a polynomial of order $N_b$ with respect to $\mu$;
\begin{equation}
\res_{\mu=\la}f(x,\la,\mu)=R(\la);\label{new9} 
\end{equation} 
and for some $\No\times{N_b}$ (see (\ref{Nnanb})) constant matrix $C$ obeying Condition~\ref{cC} we have that
\begin{equation}
\sum_{m=1}^{\No}F(x,\la,\ka_m)C_{mk}=0,\quad k=1,\ldots,N_b.\label{f42}
\end{equation} 
\end{definition}
Then we prove that
\begin{theorem}\label{th1}
Function $(\la-\mu)F(x,\la,\mu)$ is an entire function of $\la$ for any $x$ such that $\tau'(x)\neq0$ 
(see (\ref{tau})) and any $\mu\in\CC$, product $(\la-\mu)f(x,\la,\mu)$ is a polynomial of order 
$N_a$ with respect to $\la$. Values of the function $F(x,\la,\mu)$ at points $\la=\ka_{n}$ obey
\begin{equation}
\sum_{l=1}^{\No}C'_{jl}F(x,\ka_l,\mu)=0,\quad j=1,\ldots,N_a,\label{f41}
\end{equation}
where $C'$ is matrix defined in (\ref{d12}). Function $f(x,\la,\mu)$ has representation
\begin{align}
f(x,\la,\mu)&=\dfrac{R(\mu)}{\mu-\la}-R(\mu)\sum_{j=1}^{N_a}\la^{j-1}\sum_{m=1}^{\No}\bigl((C'e^{-\ell{x}}\Vo')^{-1}C'e^{-\ell{x}}\bigr)_{jm}
\dfrac{1}{\mu-\ka_{m}}=\label{new10}\\
&=\dfrac{R(\la)}{\mu-\la}+R(\la)\sum_{n=1}^{\No}\dfrac{1}{\la-\ka_{n}}\sum_{j=1}^{N_b}\bigl(e^{\ell{x}}C(\Vo{e^{\ell{x}}}C)^{-1}\bigr)_{nj}
\mu^{j-1}.\label{new11}
\end{align}
\end{theorem}

\textsl{Proof.\/} Taking into account that by condition product $(\mu-\la)f(x,\la,\mu)$ is a polynomial with respect to 
$\mu$ we can write
\begin{equation}
(\mu-\la)f(x,\la,\mu)=\sum_{m=1}^{\No}f(x,\la,\ka_{m})\dfrac{(\ka_{m}-\la)R(\mu)}{(\mu-\ka_{m})r_{m}
},\label{1th}
\end{equation}
where we used (\ref{Rla}) and (\ref{Rr}). Then by (\ref{new9}) we get
\begin{equation}
\sum_{m=1}^{\No}\dfrac{f(x,\la,\ka_{m})}{r_{m}}=-1.\label{1th0}
\end{equation}
On the other side, polynomial in (\ref{1th}) is of order $\No-1$, while by condition it must be of the order $N_b<\No$ (cf.\ 
(\ref{Nnanb})). Thus by 
(\ref{Rla5}) (with $\la$ substituted by $\mu$) we have that
\begin{equation*}
\sum_{m=1}^{\No}f(x,\la,\ka_{m})\dfrac{(\ka_{m}-\la)\ka_{m}^{j}}{r_{m}}=0,\qquad0\leq{j}\leq{N}_a-2,
\end{equation*}
where it is assumed that $N_a\geq2$. Thanks to (\ref{1th0}) this means that
\begin{equation*}
\sum_{m=1}^{\No}f(x,\la,\ka_{m})\dfrac{\ka_{m}^{j}}{r_{m}}=-\la^{j},\qquad0\leq{j}\leq{N}_a-1,
\end{equation*}
that is valid also for $N_a=1$. By means of notation (\ref{V}) this can be written in the matrix 
form as
\begin{equation}
\sum_{m=1}^{\No}f(x,\la,\ka_{m})(r^{-1}\Vo')_{mn}=-\la^{n-1},\quad1\leq{n}\leq{N}_a.\label{1th1}
\end{equation}
Next, Eq.\ (\ref{f42}) in terms of function $f$ means that
\begin{equation*}
\sum_{m=1}^{\No}f(x,\la,\ka_m)(e^{\ell{x}}C)_{mn}=0,\quad n=1,\ldots,N_b,
\end{equation*} 
where (\ref{eK}) was used. Thus due to (\ref{new6}) there exists such $N_a$-vector 
$\t{f}_j(x,\la)$ that
\begin{equation*}
f(x,\la,\ka_{m})=\sum_{j=1}^{N_a}\t{f}_j(x,\la)\bigl(C're^{-\ell{x}}\bigr)_{jm},
\end{equation*}
where $C'$ is a constant $N_a\times\No$ matrix obeying  (\ref{d12}). Substituting $f(x,\la,\ka_{m})$ in (\ref{1th1}) by the 
above equality and taking that $\det\tau'(x)\neq0$ into account we get
\begin{equation*}
\t{f}_j(x,\la)=-\sum_{n=1}^{N_a}\la^{n-1}\bigl(C'e^{-\ell{x}}\Vo'\bigr)^{-1}_{nj}.
\end{equation*}
Inserting this equality in the previous one we have
\begin{equation}
f(x,\la,\ka_{m})=-\sum_{j=1}^{N_a}\la^{j-1}\bigl((C'e^{-\ell{x}}\Vo')^{-1}C're^{-\ell{x}}\bigr)_{jm}, \label{1th2} 
\end{equation}
Because of (\ref{1th0}) Eq.\ (\ref{1th}) can be written in the form
\begin{equation*}
f(x,\la,\mu)=\dfrac{R(\mu)}{\la-\mu}\sum_{m=1}^{\No}\dfrac{f(x,\la,\ka_{m})}{r_ {m}}+ 
\sum_{m=1}^{\No}f(x,\la,\ka_{m})\dfrac{R(\mu)}{(\mu-\ka_{m} )r_{m}}. 
\end{equation*}
that gives (\ref{new10}) due to (\ref{1th0}). 

This proves that $(\la-\mu)F(x,\la,\mu)$ is an entire function of $\mu$ for any $x$ 
($\det\tau'(x)\neq0$) and any $\la\in\CC$. Moreover, (\ref{new10}) shows that product 
$(\la-\mu)f(x,\la,\mu)$ is polynomial of the order $N_a$ and
\begin{equation*}
\res_{\la=\mu}f(x,\la,\mu)=-R(\mu).\label{new91}
\end{equation*}
In order to prove (\ref{f41}) we notice that by (\ref{new10}) 
\begin{equation}
f(x,\ka_{l},\mu)=\dfrac{R(\mu)}{\mu-\ka_{l}}-R(\mu)\sum_{m=1}^{\No}
\bigl(\Vo'(C'e^{-\ell{x}}\Vo')^{-1}C'e^{-\ell{x}}\bigr)_{lm}\dfrac{1}{\mu-\ka_{m}},\label{new12} 
\end{equation}
where (\ref{V}) was used. Then direct summation gives: $\sum_{l=1}^{\No}C'_{jl}e^{-\ell_lx}f(x,\ka_{l},\mu)=0$, that thanks to (\ref{new8}) is 
just (\ref{f41}). We have also an analog of Eq.\ (\ref{1th1}) that sounds as
\begin{equation}
\sum_{l=1}^{\No}(\Vo{r^{-1}})_{ml}f(x,\ka_{l},\mu)=\mu^{m-1},\quad1\leq{m}\leq{N_{b}}.\label{1th3}
\end{equation}
This follows if in the first term of (\ref{new12}) we use (\ref{Rr}), while the second term cancels out thanks to 
(\ref{d12'}). 

We see that the properties of the function $f(x,\la,\mu)$ are symmetric with respect to the variables $\la$ and $\mu$. Thus it is natural to expect 
that besides (\ref{new10}) it also has representation  (\ref{new11}). Indeed, it is clear that this representation gives correctly 
polynomial structure of $(\mu-\la)f(x,\la,\mu)$. Thus validity of (\ref{new11}) is equivalent to condition that values 
\begin{equation}
f(x,\la,\ka_m)=-\dfrac{R(\la)}{\la-\ka_m}+R(\la)\sum_{l=1}^{\No}\dfrac{1}{\la-\ka_{l}}\bigl(e^{\ell{x}}C(\Vo{e^{\ell{x}}}C)^{-1}\Vo\bigr)_{lm}
\label{1th4}
\end{equation}
 obey (\ref{f42}). But this follows by the the direct summation and Eq.\ (\ref{new8}). Notice that the above relation is symmetric to 
(\ref{new12}). Let us mention, that (\ref{1th4}) gives (\ref{1th1}) by the same arguments as (\ref{1th3}) was derived. $\blacksquare$
\begin{corollary}\label{cor}
Following proof of this theorem one can exchange $\la$ and $\mu$,  matrices $C$ and $C'$, relations (\ref{f42}) and  
(\ref{f41}), etc.\  in Definition \ref{def1} and in Theorem~\ref{th1}.
\end{corollary}
\begin{remark}\label{rem2}
 Relations (\ref{1th2}), (\ref{new12}) and (\ref{1th4}) shows that values $f(x,\ka_{l},\mu)$ and $f(x,\la,\ka_{m})$ are 
regular functions of $\la$ and $\mu$ thanks to (\ref{Rla}) in spite of the pole behavior of $f(x,\la,\mu)$ at $\la=\mu$. 
Nevertheless, the pole behavior of $f(x,\la,\mu)$ has essential consequence for properties of these values. Thanks to 
(\ref{1th2}) and (\ref{V}) we have that 
$f(x,\la,\ka_{m})|_{\la=\ka_{l}}=-\bigl(\Vo'(C'e^{-\ell{x}}\Vo')^{-1}C're^{-\ell{x}}\bigr)$. On the other side by 
(\ref{new12}) and (\ref{Rla}) 
$f(x,\ka_{l},\mu)|_{\mu=\ka_{m}}=r_{m}\delta_{l,m}-\bigl(\Vo'(C'e^{-\ell{x}}\Vo')^{-1}C're^{-\ell{x}}\bigr)$, so that
\begin{equation}
 F(x,\ka_{l},\mu)|_{\mu=\ka_{m}}=r_m\delta_{l,m}+F(x,\la,\ka_{m})|_{\la=\ka_{l}},\label{1th5}
\end{equation}
where (\ref{new8}) was used.

\end{remark}

\section{Properties of the Cauchy--Jost function}\label{properties}

Relations (\ref{new10}) and (\ref{new11}) can be written in the form
\begin{align*}
\tau'(x)f(x,\la,\mu)&=\dfrac{R(\mu)\tau'(x)}{\mu-\la}-\\
&-R(\mu)\sum_{m=1}^{\No}\sum_{i,j=1}^{N_a}(-1)^{i+j}\la^{j-1}
\det\bigl(C'_{\h{i}}{e}^{-\ell{x}}\Vo'_{\h{j}}\bigr)C'_{im}\dfrac{e^{-\ell_mx}}{\mu-\ka_{m}},\\
\tau(x)f(x,\la,\mu)&=\dfrac{R(\la)\tau(x)}{\mu-\la}+R(\la)\sum_{n=1}^{\No}\dfrac{e^{\ell_nx}}{\la-\ka_{n}}\sum_{i,j=1}^{N_b}(-1)^{i+j}C_{ni}
\det\bigl(\Vo_{\h{j}}{e}^{\ell{x}}C_{\h{i}}\bigr)\mu^{ j-1},
\end{align*}
that follows from (\ref{tau}) and where notation $\Vo_{\h{ij}}$ denotes matrix with removed $j$-th 
row and $i$-th column.  Let also $C'_{\h{i}}$ and $\Vo_{\h{i}}$ denote matrices $C'$ and $\Vo$ with removed $i$-th row 
and $C_{\h{i}}$ and $\Vo'_{\h{i}}$ denote matrices $C$ and $\Vo'$ with removed $i$-th column. Now using the Binet--Cauchy formula 
like in (\ref{tauf1}), (\ref{tauf2}) we get
 \begin{align*}
\tau'(x)f(x,\la,\mu)&=\dfrac{R(\mu)\tau'(x)}{\mu-\la}-R(\mu)\sum_{m=1}^{\No}\dfrac{e^{-\ell_mx}}{\mu-\ka_{m}}\sum_{i,j=1}^{N_a}(-1)^{i+j}\la^{j-1}
C'_{im}\times\\
&\times\sum_{1\leq{n_1<\ldots<n_{N_a-1}\leq\No}}
C'_{\h{i}}(n_1,\ldots,n_{N_a-1})\Vo'_{\h{j}}(n_1,\ldots,n_{N_a-1})\prod_{k=1}^{N_a-1}{e}^{-\ell_{n_k}{x}},\\
\tau(x)f(x,\la,\mu)&=\dfrac{R(\la)\tau(x)}{\mu-\la}+R(\la)\sum_{n=1}^{\No}
\dfrac{e^{\ell_nx}}{\la-\ka_{n}}\sum_{i,j=1}^{N_b}(-1)^{i+j}C_{ni}\mu^{j-1}\times\\
&\times\sum_{1\leq{n_1<\ldots<n_{N_b-1}\leq\No}}C_{\h{i}}(n_1,\ldots,n_{N_b-1})\Vo_{\h{j}}(n_1,\ldots,n_{N_b-1})
\prod_{k=1}^{N_b-1}{e}^{\ell_{n_k}{x}}.
 \end{align*}
Taking the standard relations of the kind 
\begin{align*}
&\sum_{i=1}^{N_b}(-1)^{i}C_{mi}C_{\h{i}}(n_1,\ldots,n_{N_b-1})=-C(m,n_1,\ldots,n_{N_b-1}),\\
&\sum_{j=1}^{N_a}(-1)^{j}\la^{j-1}\Vo'_{\h{j}}(n_1,\ldots,n_{N_a-1})=-\prod_{j=1}^{N_a-1}(\ka_{n_j}-\la)\Vo'(n_1,\ldots,n_{N_a-1})
\end{align*}
into account we rewrite the above equalities in the form
 \begin{align*}
&\tau'(x)f(x,\la,\mu)=\dfrac{R(\mu)\tau'(x)}{\mu-\la}-R(\mu)\sum_{m=1}^{\No}\dfrac{e^{-\ell_mx}}{\mu-\ka_{m}}\times\\
&\qquad\times\sum_{1\leq{n_1<\ldots<n_{N_a-1}\leq\No}}
C'(m,n_1,\ldots,n_{N_a-1})\Vo'(n_1,\ldots,n_{N_a-1})\prod_{k=1}^{N_a-1}{e}^{-\ell_{n_k}{x}}(\ka_{n_k}-\la),\\
&\tau(x)f(x,\la,\mu)=\dfrac{R(\la)\tau(x)}{\mu-\la}+R(\la)\sum_{n=1}^{\No}\dfrac{e^{\ell_nx}}{\la-\ka_{n}}\times\\
&\qquad\times\sum_{1\leq{n_1<\ldots<n_{N_b-1}\leq\No}}C(n,n_1,\ldots,n_{N_b-1})\Vo(n_1,\ldots,n_{N_b-1})
\prod_{k=1}^{N_b-1}{e}^{\ell_{n_k}{x}}(\ka_{n_k}-\mu).
\end{align*}
We introduce two new functions, $\tau(x,\la,\mu)$ and  $\tau'(x,\la,\mu)$ in such a way that
\begin{equation}
f(x,\la,\mu)=\dfrac{\tau(x,\la,\mu)}{(\mu-\la)\tau(x)}=\dfrac{\tau'(x,\la,\mu)}{(\mu-\la)\tau'(x)},\label{new13} 
\end{equation}
so that due to the above 
\begin{align*}
&\dfrac{\tau'(x,\la,\mu)}{R(\mu)}=\tau'(x)-(\mu-\la)\sum_{m=1}^{\No}\dfrac{e^{-\ell_mx}}{\mu-\ka_{m}}\times\nonumber\\
&\qquad\times\sum_{1\leq{n_1<\ldots<n_{N_a-1}\leq\No}}
C'(m,n_1,\ldots,n_{N_a-1})\Vo'(n_1,\ldots,n_{N_a-1})\prod_{k=1}^{N_a-1}{e}^{-\ell_{n_k}{x}}(\ka_{n_k}-\la),\\
&\dfrac{\tau(x,\la,\mu)}{R(\la)}=\tau(x)+(\mu-\la)\sum_{n=1}^{\No}\dfrac{e^{\ell_nx}}{\la-\ka_{n}}\times\nonumber\\
&\qquad\times\sum_{1\leq{n_1<\ldots<n_{N_b-1}\leq\No}}C(n,n_1,\ldots,n_{N_b-1})\Vo(n_1,\ldots,n_{N_b-1})
\prod_{k=1}^{N_b-1}{e}^{\ell_{n_k}{x}}(\ka_{n_k}-\mu).
 \end{align*}
Thanks to (\ref{new13}) and Definition \ref{def1} ratio  $\dfrac{\tau'(x,\la,\mu)}{R(\mu)}$ decays as $\mu^{-N_a}$ when
$\mu\to\infty$, and thanks to the Theorem~\ref{th1} ratio  $\dfrac{\tau(x,\la,\mu)}{R(\la)}$ decays as $\la^{-N_b}$ when  
$\la\to\infty$. Thus
\begin{align}
&\dfrac{\tau(x,\la,\mu)}{R(\la)}=\sum_{1\leq{n}_1<\ldots<n_{N_b}\leq\No}C(\{n_{i}\})\Vo(\{n_{i}\})\prod_{j=1}^{N_{b}}e^{\ell_{n_{j}}x}_
{}\dfrac{\ka_{n_j}-\mu}{\ka_{n_j}-\la},\label{new151}\\
&\dfrac{\tau'(x,\la,\mu)}{R(\mu)}=\sum_{1\leq{n}_1<\ldots<n_{N_a}\leq\No}C'(\{n_{i}\})\Vo'(\{n_{i}\})\prod_{j=1}^{N_{a}}e^{-\ell_{n_{j}}x}_{} 
\dfrac{\ka_{n_j}-\la}{\ka_{n_j}-\mu},\label{new152}
\end{align}
as residuals of the r.h.s.'s of these equalities coincide with residuals of the previous ones thanks to (\ref{Vo}) and (\ref{Rla}). Notice 
that these relations follow from expressions for $\tau(x)$ and $\tau'(x)$ in (\ref{tauf1}) and (\ref{tauf2}) under 
substitution $e^{\ell_{n}x}\to e^{\ell_{n}x}\dfrac{\ka_{n}-\mu}{\ka_{n}-\la}$, that up to some unessential constants is 
nothing but a double Miwa shift in (\ref{new2}). Finally we get
\begin{align}
&\tau(x,\la,\mu)=\sum_{1\leq{n}_1<\ldots<n_{N_b}\leq\No}C(\{n_{i}\})\Vo(\{n_{i}\})\prod_{\{n_i\}}e^{\ell_{n_{i}}x}_{}
(\mu-\ka_{n_i})\prod_{\{\bar{n}_j\}}(\la-\ka_{\bar{n}_j}),\label{new15}\\
&\tau'(x,\la,\mu)=\sum_{1\leq{n}_1<\ldots<n_{N_a}\leq\No}C'(\{n_{i}\})\Vo'(\{n_{i}\})\prod_{\{n_i\}}e^{-\ell_{n_{i}}x}_{} 
(\la-\ka_{n_i})\prod_{\{\bar{n}_j\}}(\mu-\ka_{\bar{n}_j}),\label{new16}
\end{align}
where $\{n_{i}\}$ denotes set of indexes involved in summation and $\{\bar{n}_j\}$ denotes set that is complimentary to 
$\{n_{i}\}$ in $\{1,\ldots,\No\}$. 

In virtue of  (\ref{Rla}) and   (\ref{tauf1}),   (\ref{tauf2}) relations (\ref{new15}) and (\ref{new16}) prove
\begin{equation}
\tau(x,\la,\la)=R(\la)\tau(x),\qquad \tau'(x,\la,\la)=R(\la)\tau'(x).\label{new17}
\end{equation}
The Jost solution $\vhi(x,\mu)$ of the heat equation, Jost solution $\psi(x,\la)$ of the dual heat equation and the potential $u(x)$ of this 
equation are given as 
\begin{align}
&\vhi(x,\mu)=-\lim_{\la\to\infty}\la^{-N_a+1}f(x,\la,\mu)e^{\ell(\mu)x},\label{new18}\\
&\psi(x,\la)=\lim_{\mu\to\infty}\mu^{-N_b+1}f(x,\la,\mu)e^{-\ell(\la)x},\label{new19}\\
&u(x)=- 2\lim_{\la,\mu\to\infty}\la^{-N_a+1}\mu^{-N_b+1}f_{x_1}(x,\la,\mu),\label{new20}
\end{align}
so that thanks to (\ref{new13})--(\ref{new16}):
\begin{align}
&\vhi(x,\mu)=\dfrac{e^{\ell(\mu)x}}{\tau(x)}\sum_{1\leq{n}_1<\ldots<n_{N_b}\leq\No}C(\{n_{i}\})\Vo(\{n_{i}\})\prod_{\{n_i\}}e^{\ell_{n_{i}}x}_{}
(\mu-\ka_{n_i}),\label{phi}\\
&\psi(x,\la)=\dfrac{e^{-\ell(\la)x}}{\tau'(x)}\sum_{1\leq{n}_1<\ldots<n_{N_a}\leq\No}C'(\{n_{i}\})\Vo'(\{n_{i}\})\prod_{\{n_i\}}e^{-\ell_{n_{i}}x}_{} 
(\la-\ka_{n_i}),\label{psi}
\end{align}
while for the potential $u(x)$ we rederive (\ref{u}). It is worth to mention that in the pure solitonic case the Jost 
solutions themselves can be defined by means of the obvious analogs of the Definition \ref{def1}. In order to avoid 
singularities with respect to the spectral parameter we normalize the Jost and dual Jost solutions in a way that 
$\vhi(x,\mu)e^{-\ell(\mu)x}$ is a polynomial of order $N_b$ with respect to $\mu$ and $\psi(x,\la)e^{\ell(\la)x}$ is 
a polynomial of order $N_a$ with respect to $\la$.

In order to get evolutions of the Cauchy--Jost function with respect to times $x_1,x_2,\ldots$ we differentiate 
(\ref{new10}):
\begin{align*}
 f_{x_k}(x,\la,\mu)&=-R(\mu)\sum_{j=1}^{N_a}\la^{j-1}\sum_{m=1}^{\No}\biggl\{\sum_{n=1}^{\No}\bigl((C'e^{-\ell 
x}\Vo')^{-1}C'e^{-\ell x}\bigr)_{j,n}\ka_{n}^{k}\times\\
&\times\bigl(\Vo'(C'e^{-\ell x}\Vo')^{-1}C'e^{-\ell x}\bigr)_{n,m}-\bigl((C'e^{-\ell 
x}\Vo')^{-1}C'e^{-\ell x}\bigr)_{j,m}\biggr\}\dfrac{1}{\mu-\ka_{m}},
\end{align*}
where (\ref{V}) was used. By means of (\ref{1th2}) and (\ref{new10}) this can be written as
\begin{equation}
 f_{x_k}(x,\la,\mu) =-\sum_{n=1}^{\No}f(x,\la,\ka_{n})\dfrac{\ka_{n}^{k}}{r_{n}} f(x,\ka_{n},\mu). \label{new21}
\end{equation}
Taking analyticity properties of $f(x,\la,\mu)$ into account, we have
\begin{equation*}
 f_{x_k}(x,\la,\mu) =f(x,\la,\mu)(\la^{k}-\mu^{k})-\dfrac{1}{2\pi 
i}\oint_{\gamma}\dfrac{d\nu\,\nu^{k}}{R(\nu)}f(x,\la,\nu)f(x,\nu,\mu),
\end{equation*}
where we used (\ref{Rr}) and where contour $\gamma$ encircles all $\ka_1,\ldots,\ka_{\No}$, $\la$ and $\mu$. Because of 
(\ref{ellx}) and (\ref{new8}) this means
\begin{equation}
F_{x_k}(x,\la,\mu) =\dfrac{-1}{2\pi 
i}\oint_{\gamma}\dfrac{d\nu\,\nu^{k}}{R(\nu)}F(x,\la,\nu)F(x,\nu,\mu), \label{new22}
\end{equation}
In particular, thanks to 
(\ref{Nnanb}), (\ref{Rla}) and (\ref{new18}), (\ref{new19}) we have
\begin{equation}
F_{x_1}(x,\la,\mu) =\psi(x,\la)\vhi(x,\mu),\label{new23}
\end{equation}
that gives (\ref{new1}). 

\section{Darboux transformations}\label{Darboux}

We start with the ($N_a,N_b$)-soliton potential and denote by tilde all Darboux transformed objects. Taking into account 
that the soliton solutions are parametrized by two discrete parameters, $N_a$ and $N_b$, we can classify three types of 
Darboux transformations: $A$-transformation, that changes $N_a\to{N_a}+1$ with $N_b$ being fixed, $B$-traansformation, 
that preserves $N_a$ and changes $N_b\to{N_b}+1$ and 0-transformations, that change only matrices $C$ and $C'$, but not $N_a$ 
and $N_b$. Here we also consider inverse transformations, $A^{-1}$ and $B^{-1}$. For $A$- and $B$-types we denote the
added parameter $\ka$ as $\ka_{\No+1}$ and, correspondingly,
\begin{equation}
\t{R}(\la)=\prod_{n=1}^{\No+1}(\la-\ka_{n})=(\la-\ka_{\No+1})R(\la),\label{tA} 
\end{equation} 
see (\ref{Rla}). Then (cf.\ (\ref{r}) and (\ref{Rr}))
\begin{equation}
\t{r}_{n}=(\ka_{n}-\ka_{\No+1})r_n,\quad{n}=1,\ldots,\No,\qquad\t{r}_{\No+1}=R(\ka_{\No+1}).\label{o108} 
\end{equation} 
In the case of $0$-transformation $\t{R}(\la)=R(\la)$. 

In \cite{BPP2015} we proved the specific property of the Cauchy--Jost function: its time evolutions with respect to KPII hierarchy are given in 
terms of the Cauchy--Jost function itself. Here we prove that Darboux transformation in terms of this function is given by 
means of linear equations.
\begin{theorem}
Darboux transformed Cauchy--Jost function $\t{F}(x,\la,\mu)$ is given as solution of the following equations:
\begin{align}
 &\text{A-transformation: }\dfrac{\t{F}(x,\la,\mu)}{\t{R}(\la)}=\dfrac{F(x,\la,\mu)}{R(\la)}-\dfrac{F(x,\la,\ka_{\No+1})}{R(\la)}
\dfrac{\t{F}(x,\ka_{\No+1},\mu)}{\t{r}_{\No+1}}, \label{o-549}\\
&\text{and auxiliary condition: }\sum_{l=1}^{\No+1}\dfrac{a_{l}}{\t{r}_{l}}\t{F}(x,\ka_{l},\mu)=0,\label{o-551} 
\end{align}
where $a_1,\ldots,a_{\No+1}$ are arbitrary real parameters, $a_{\No+1}\neq0$; 
\begin{align}
&\text{B-transformation: }\dfrac{\t{F}(x,\la,\mu)}{\t{R}(\mu)}=\dfrac{F(x,\la,\mu)}{R(\mu)}+\dfrac{\t{F}(x,\la,\ka_{\No+1})}{\t{r}_{\No+1}}
\dfrac{F(x,\ka_{\No+1},\mu)}{R(\mu)},\label{o-66}\\
&\text{and auxiliary condition: } 
\sum_{m=1}^{\No+1}\t{F}(x,\la,\ka_{m})\dfrac{b_m}{\t{r}_{m}}=0,\label{o45}
\end{align}
where $b_1,\ldots,b_{\No+1}$ are arbitrary real parameters, $b_{\No+1}\neq0$;
and
\begin{align}
\text{0-transformation: }
\t{F}(x,\la,\mu)&=F(x,\la,\mu)-\nonumber\\
&-\sum_{i=1}^{\No}\t{F}(x,\la,\ka_{i})b_i\sum_{j=1}^{\No}a_jF(x,\ka_{j},\mu),\label{01}
\end{align}
where $a_1,\ldots,a_{\No}$ and $b_1,\ldots,b_{\No}$ are arbitrary real parameters such that 
\begin{equation}
(arb)\equiv\sum_{n=1}^{\No}a_{n}r_{n}b_{n}\neq-1.\label{02} 
\end{equation}
\end{theorem}
\begin{remark}
 Cauchy--Jost functions constructed by these means can have singularities with respect to $x$-variables. We discuss this 
problem below.
\end{remark}

\textsl{Proof for the $A$-transformation.\/} Let $\t{f}$ be defined by $\t{F}$ as in (\ref{new8}). By (\ref{tA}) we 
can rewrite (\ref{o-549}) as
\begin{equation}
\t{f}(x,\la,\mu)=(\la-\ka_{\No+1})\biggl\{f(x,\la,\mu)-\dfrac{f(x,\la,\ka_{\No+1})}{R(\ka_{\No+1})}\t{f}(x,\ka_{\No+1}, 
\mu)\biggr\},\label{o-55}
\end{equation} 
that proves (\ref{new9}) for the transformed $\t{f}$ and proves it to be a polynomial of order $N_a+1$ with respect to 
$\la$, if $\t{f}$ exists. Notice that in virtue of (\ref{new9}) the value $\t{f}(x,\ka_{\No+1},\mu)$ is not defined by this 
equality as for $\la=\ka_{\No+1}$ it gives identity. This remark motivates condition on $a_{\No+1}$ to be different from 
zero. Reduction of (\ref{o-549}) gives
\begin{equation}
\dfrac{\t{F}(x,\ka_{l},\mu)}{\t{r}_{l}}=\dfrac{F(x,\ka_{l},\mu)}{r_{l}}-\dfrac{F(x,\ka_{l},\ka_{\No+1})}{r_{l}}\dfrac{\t{F}
(x,\ka_{\No+1},\mu)}{\t{r}_{\No+1}},\quad l=1,\ldots,\No,\label{o-552}
\end{equation}
By means of (\ref{f41}) we get from (\ref{o-552}) that
\begin{equation}
\sum_{l=1}^{\No}C'_{jl}\dfrac{r_{l}}{\t{r}_{l}}\t{F}(x,\ka_{l},\mu)=0,\quad j=1,\ldots,N_a.\label{o-57} 
\end{equation} 
Matrix $\t{C}'$ corresponding to the transformed case must be of size $(N_a+1)\times{(\No+1)}$, so we use (\ref{o-551}) to 
sum up (\ref{o-552}) from 1 to $\No$:
\begin{equation*}
a_{\No+1}\dfrac{\t{F}(x,\ka_{\No+1},\mu)}{\t{r}_{\No+1}}=-\sum_{l=1}^{\No}a_{l}\dfrac{F(x,\ka_{l},\mu)}{r_{l}}+\dfrac{\t{F}(x
,\ka_{\No+1},\mu)}{\t{r}_{\No+1}}
\sum_{l=1}^{\No}a_{l}\dfrac{F(x,\ka_{l},\ka_{\No+1})}{r_{l}},
\end{equation*}
that defines
\begin{equation}
\dfrac{\t{F}(x,\ka_{\No+1},\mu)}{\t{r}_{\No+1}}=\dfrac{\displaystyle\sum_{l=1}^{\No}\dfrac{a_l}{r_{l}}F(x,\ka_{l},\mu)}
{\displaystyle\sum_{l=1}^{\No}\dfrac{a_l}{r_{l}}F(x,\ka_{l},\ka_{\No+1})-a_{\No+1}}.\label{o-58} 
\end{equation} 
Inserting this in the r.h.s.\ of (\ref{o-549}) we get  
\begin{equation}
\t{F}(x,\la,\mu)=(\la-\ka_{\No+1})\left\{F(x,\la,\mu)-\dfrac{F(x,\la,\ka_{\No+1})\displaystyle\sum_{l=1}^{\No}\dfrac{a_l}{r_
{ l}}F(x, \ka_ {l} , \mu) }
{\displaystyle\sum_{l=1}^{\No}\dfrac{a_l}{r_{l}}F(x,\ka_{l},\ka_{\No+1})-a_{\No+1}}\right\},\label{o-1}
\end{equation} 
that has the standard form of the binary Darboux transformation as consequence: we differentiate it by $x_1$ and 
factorize with respect to $\la$ and $\mu$ by (\ref{new23}). The result specialize the transformations introduced in 
\cite{MS}.

In order to complete the proof we have to show that there exists matrix $\t{C}'$ that obeys tilde version of relation 
(\ref{f41}). This matrix is given, in fact, by equality (\ref{o-57}) and auxiliary condition (\ref{o-551}). Expression 
for the matrix $\t{C}$ obeying tilde analog of (\ref{d12}), $\t{C}'\t{r}\t{C}=0$, follows by (\ref{o-1}) and (\ref{f42}). 
Summarizing we get
\begin{equation}
\t{C}=\left(\begin{array}{c}
C\\ \dfrac{-aC}{a_{\No+1}}\end{array}\right),\qquad
\t{C}'\t{r}=\left(\begin{array}{cc}
C'r&\bf{0}\\
a&a_{\No+1}\end{array}\right),\label{o-615} 
\end{equation} 
where we introduced row
\begin{equation}
a=(a_1,\ldots,a_{\No}),\label{a} 
\end{equation} 
and where $\bf{0}$ denotes zero $\No$-row. 

\textsl{Proof for the $B$-transformation} is close to the above.  By (\ref{new8}) and (\ref{tA}) we 
rewrite (\ref{o-66}) as
\begin{equation}
\t{f}(x,\la,\mu)=(\mu-\ka_{\No+1})\biggl\{f(x,\la,\mu)+\dfrac{\t{f}(x,\la,\ka_{\No+1})}{R(\ka_{\No+1})}f(x,\ka_{\No+1},
\mu)\biggr\},\label{o-5511}
\end{equation} 
that proves (\ref{new9}) for the transformed $\t{f}$ and proves it to be a polynomial of order $N_b+1$ with respect to 
$\mu$. Value $\t{f}(x,\la,\ka_{\No+1})$ is not defined by this equality, so we use (\ref{o45}) under condition 
$b_{\No+1}\neq0$. Taking into account that for $m=1,\ldots,\No$ reduction of (\ref{o-66}) gives
\begin{equation}
\dfrac{\t{F}(x,\la,\ka_{m})}{\t{r}_{m}}=\dfrac{F(x,\la,\ka_{m})}{r_{m}}+\dfrac{\t{F}(x,\la,\ka_{\No+1})}{\t{r}_{\No+1}}
\dfrac{F(x,\ka_{\No+1},\ka_{m})}{r_{m}},\label{o-553}
\end{equation}
we get thanks to (\ref{o45})
\begin{equation}
\dfrac{\t{F}(x,\la,\ka_{\No+1})}{\t{r}_{\No+1}}=-\dfrac{\displaystyle\sum_{m=1}^{\No}F(x,\la,\ka_{m})\dfrac{b_m}{r_{m}}}
{\displaystyle\sum_{m=1}^{\No}F(x,\ka_{\No+1},\ka_{m})\dfrac{b_m}{r_{m}}+b_{\No+1}},\label{o-581} 
\end{equation} 
so that finally
\begin{equation}
\t{F}(x,\la,\mu)=(\mu-\ka_{\No+1})\left\{F(x,\la,\mu)-\dfrac{F(x,\ka_{\No+1},\mu)\displaystyle\sum_{m=l}^{\No}
F(x, \la,\ka_ {m})\dfrac{b_m}{r_{m}}}
{\displaystyle\sum_{m=1}^{\No}F(x,\ka_{\No+1},\ka_{m})\dfrac{b_m}{r_{m}}+b_{\No+1}}\right\}.\label{o-11}
\end{equation} 
This equality leads to the standard form of the binary Darboux transformation,  \cite{MS}. We differentiate (\ref{o-11})
by $x_1$ and factorize with respect to $\la$ and $\mu$ dependence by (\ref{new23}).

In order to complete the proof we have to show that there exist matrices $\t{C}$ and $\t{C}'$ that obey tilde versions of 
relations (\ref{f42}) and (\ref{f41}). By means of (\ref{f42}) we get from (\ref{o-553}) that
\begin{equation}
\sum_{m=1}^{\No}\t{F}(x,\la,\ka_{m})\dfrac{r_{m}}{\t{r}_{m}}C_{mj}=0,\quad j=1,\ldots,N_b,\label{o-571} 
\end{equation} 
that together with (\ref{o45}) shows that these matrices obeying analog of (\ref{d12}) are
\begin{equation}
\t{r}\t{C}=\left(\begin{array}{cc}
rC&b\\ \mathbf{0}&b_{\No+1}\end{array}\right),\qquad
\t{C}'=\biggl(C', \dfrac{-C'b}{b_{\No+1}}\biggr),\label{o-6151} 
\end{equation} 
where we introduced column
\begin{equation}
b=(b_1,\ldots,b_{\No})^{\text{T}},\label{b} 
\end{equation} 
and where $\bf{0}$ denotes zero $\No$-row. 

\textsl{Proof for the $0$-type transformation.} Let us denote
\begin{align}
 &F(x,\ka_{l},\ka_{m})=F(x,\ka_{l},\mu)\bigr|_{\mu=\ka_{m}},\label{03}\\
 \intertext{so that by (\ref{1th5})}
 &F(x,\la,\ka_{m})\bigr|_{\la=\ka_{l}}=F(x,\ka_{l},\ka_{m})-r_{l}\delta_{l,m}.\label{04}
\end{align}
Then setting $\mu=\ka_{m}$ in (\ref{01}), multiplying it by $a_m$ and summing up we get
\begin{equation}
\sum_{m=1}^{\No}\t{F}(x,\la,\ka_{m})b_m=
\dfrac{\sum_{i=1}^{\No}F(x,\la,\ka_{i})b_i}{\sum_{i,j=1}^{\No}a_jF(x,\ka_{j},\ka_{i})b_i+1 },\label{04'}
\end{equation}
so that by (\ref{01}) we derive
\begin{align}
\t{F}(x,\la,\mu)&=F(x,\la,\mu)-\nonumber\\
&-\dfrac{\sum_{i=1}^{\No}F(x,\la,\ka_{i})b_i\sum_{j=1}^{\No}a_jF(x,\ka_{j},\mu)}
{\sum_{i,j=1}^{\No}a_jF(x,\ka_{j},\ka_{i})b_i+1}.\label{05}
\end{align}
Properties of  $\t{F}(x,\la,\mu)$ with respect to $\la$ and $\mu$ are determined by the properties of 
$F(x,\la,\mu)$ and obviously coincide with them. By reductions of (\ref{05}) and thanks to   (\ref{03}) and (\ref{04}) we get
\begin{align}
\t{F}(x,\ka_{l},\mu)&=F(x,\ka_{l},\mu)-\nonumber\\
&-\dfrac{\sum_{i=1}^{\No}(F(x,\ka_{l},\ka_{i})-r_{l}\delta_{l,i})b_i\sum_{j=1}^{\No}a_jF(x,\ka_{j},\mu)}
{\sum_{i,j=1}^{\No}a_jF(x,\ka_{j},\ka_{i})b_i+1},\label{06}\\
\t{F}(x,\ka_{l},\mu)\bigr|_{\mu=\ka_{m}}&=F(x,\ka_{l},\ka_{m})-\nonumber\\
&-\dfrac{\sum_{i=1}^{\No}(F(x,\ka_{l},\ka_{i})-r_{l}\delta_{l,i})b_i\sum_{j=1}^{\No}a_jF(x,\ka_{j},\ka_{m})}
{\sum_{i,j=1}^{\No}a_jF(x,\ka_{j},\ka_{i})b_i+1},\label{07}\\
\t{F}(x,\la,\ka_{m})&=F(x,\la,\ka_{m})-\nonumber\\
&-\dfrac{\sum_{i=1}^{\No}F(x,\la,\ka_{i})b_i\sum_{j=1}^{\No}a_jF(x,\ka_{j},\ka_{m})}
{\sum_{i,j=1}^{\No}a_jF(x,\ka_{j},\ka_{i})b_i+1},\label{08}\\
\t{F}(x,\la,\ka_{m})\bigr|_{\la=\ka_{l}}&=F(x,\la,\ka_{m})-r_{l}\delta_{l,m}-\nonumber\\
&-\dfrac{\sum_{i=1}^{\No}(F(x,\ka_{l},\ka_{i})-r_{l}\delta_{l,i})b_i\sum_{j=1}^{\No}a_jF(x,\ka_{j},\ka_{m})}
{\sum_{i,j=1}^{\No}a_jF(x,\ka_{j},\ka_{i})b_i+1}.\label{09}
\end{align}
 
Relations (\ref{07}) and (\ref{09}) proves that $\t{F}(x,\la,\mu)$ obeys 
\begin{equation*}
\t{F}(x,\ka_{l},\mu)\bigr|_{\mu=\ka_{m}}= \t{F}(x,\la,\ka_{m})\bigr|_{\la=\ka_{l}}+r_{l}\delta_{l,m},
\end{equation*}
 i.e., tilde analog of relation (\ref{1th5}) for ${F}(x,\la,\mu)$ with the same matrix $r$. Next, by (\ref{06})
 \begin{equation}
\sum_{j=1}^{\No}a_j\t{F}(x,\ka_{j},\mu)=(1+(arb))\dfrac{\sum_{j=1}^{\No}a_j{F}(x,\ka_{j},\mu)}
{\sum_{i,j=1}^{\No}a_jF(x,\ka_{j},\ka_{i})b_i+1},\label{010}
 \end{equation}
 where notation (\ref{02}) was used. On the other side, applying $C'$ from the left to (\ref{06}) we get by (\ref{f41})
 \begin{equation*}
  \sum_{l=1}^{\No}C'_{jl}\t{F}(x,\ka_{l},\mu)=\sum_{i=1}^{\No}C'_{ji}r_ib_i\dfrac{\sum_{j=1}^{\No}a_j{F}(x,\ka_{j},\mu)}
{\sum_{i,j=1}^{\No}a_jF(x,\ka_{j},\ka_{i})b_i+1},
 \end{equation*}
 that thanks to condition (\ref{02}) and (\ref{010}) proves tilde version 
of relation (\ref{f41}), where
\begin{equation}
 \t{C}'=C'\biggl(E_{\No\times\No}-\dfrac{rb\otimes{a}}{1+(arb)}\biggr)\label{011}
\end{equation}
and where we used notations (\ref{a}) and (\ref{b}) for row $a$ and column $b$. 

Next, we apply matrix $C$ to (\ref{08}) from the right. Notice that due to (\ref{03}) and (\ref{04}) equality (\ref{f42}) 
gives $\sum_{m=1}^{\No}F(x,\ka_{l},\ka_{m})C_{m,j}=r_{l}C_{l,j}$. Then like above we derive that  
$\sum_{m=1}^{\No}\t{F}(x,\la,\ka_{m})\t{C}_{m,j}=0$, i.e., tilde-version of equality (\ref{f42}) with
\begin{equation}
 \t{C}=\bigl(E_{\No\times\No}+b\otimes{ar}\bigl)C.\label{012}
\end{equation}
Thus function $\t{F}(x,\la,\mu)$ given by (\ref{01}) obeys all properties of the Cauchy--Jost function in Definition 
\ref{def1}. Taking into account that
\begin{equation}
 \biggl(E_{\No\times\No}-\dfrac{rb\otimes{a}}{1+(arb)}\biggr)r\bigl(E_{\No\times\No}+b\otimes{ar}\bigl)=r,\label{013}
\end{equation}
we see that new matrices $\t{C}$ and $\t{C}'$ obey relation (\ref{d12}): $\t{C}'r\t{C}=0$.

Thanks to (\ref{new6}) we can write column $b$ and row $a$ as
\begin{equation}
 b=Cv'+r{C'}^{\text{T}}v,\qquad a=w'C'+wC^{\text{T}}r^{-1},\label{ab}
\end{equation}
so that thanks to (\ref{d12}) $C'rb=C'r^{2}{C'}^{\text{T}}v$, $arC=wC^{\text{T}}C$ and 
$arb=w'C'r^{2}{C'}^{\text{T}}v+wC^{\text{T}}Cv'$. This means that without loss of generality one can set $w'=v'=0$. Then 
$arb=0$ and we get
\begin{equation}
\begin{split}
\t{C}=\left(E_{\No\times\No}+r{C'}^{\text{T}}v\otimes{wC^{\text{T}}}\right)C,\\
\t{C}'r=C'r\left(E_{\No\times\No}-r{C'}^{\text{T}}v\otimes{wC^{\text{T}}}\right),\label{zero} 
\end{split}
\end{equation} 
that means that 0-transformation is parametrized by $N_b$-column $v$ and $N_a$-row $w$.

\subsection{Inverse transformations}
Let now deal with the Cauchy--Jost function  $\t{F}$ parametrized by $\t{N}_{a}=N_a+1$, $\t{N}_{b}=N_b$, 
parameters 
$\ka_{1},\ldots,\ka_{\No+1}$ and by matrix $\t{C}$ (or $\t{C'}$) as (cf.\ (\ref{f42}), (\ref{f41}))
\begin{equation}\label{o-111}
\begin{split}
&\sum_{m=1}^{\No+1}\t{F}(x,\la,\ka_{m})\t{C}_{m,i}=0,\quad i=1,\ldots,N_b,\\
&\sum_{l=1}^{\No+1}\t{C}'_{j,l}\t{F}(x,\ka_{l},\mu)=0,\quad j=1,\ldots,N_a+1.
\end{split}
\end{equation}
Then the same relation (\ref{o-549}) supplies us with the transformation inverse to considered above $A$-transformation, 
i.e., gives function $F(x,\la,\mu)$ with parameters $N_a$ and $N_b$ with $\ka_{\No+1}$ omitted. First, in analogy to 
(\ref{03}) we denote 
\begin{equation}
 \t{F}(x)=\t{F}(x,\ka_{\No+1},\mu)\bigr|_{\mu=\ka_{\No+1}},\label{o-112}
\end{equation}
so thanks to Remark \ref{rem2} and the tilde-version of (\ref{1th5}) we have
\begin{equation}
\t{F}(x,\la,\ka_{\No+1})\bigr|_{\la=\ka_{\No+1}}=\t{F}(x)-\t{r}_{\No+1}.\label{o-113}
\end{equation}
Thanks to this notation equality (\ref{o-549}) at $\mu=\ka_{\No+1}$ reduces to
\begin{equation}
\dfrac{F(x,\la,\ka_{\No+1})}{R(\la)}=-\dfrac{\t{r}_{\No+1}\t{F}(x,\la,\ka_{\No+1})}{\t{R}(\la)(\t{F}(x)-r_{\No+1})},
\label{o-56 } 
\end{equation} 
so that (\ref{o-549}) gives explicitly transformed function $F(x,\la,\mu)$: 
\begin{equation}
F(x,\la,\mu)=\dfrac{1}{\la-\ka_{\No+1}}\biggl[\t{F}(x,\la,\mu)-\dfrac{\t{F}(x,\la,\ka_{\No+1})\t{F}(x,\ka_{\No+1},\mu)}
{\t{F}(x)-r_{\No+1}}\biggr],\label{o-28}
\end{equation} 

In order to find matrices $C$ and $C'$ obeying (\ref{f42}) and (\ref{f41}) notice that thanks to the first equalities in 
(\ref{o-111}) and  (\ref{o-112}) we have by (\ref{o-28})
\begin{equation}
\sum_{m=1}^{\No}F(x,\la,\ka_{m})\t{C}_{mi}=0,\label{o-29} 
 \end{equation}
for all $i=1,\ldots,N_b$. Thus $C$ equals matrix $\t{C}$ with removed last row. Next, let us consider values 
$F(x,\ka_{l},\mu)$ for $l=1,\ldots,\No$ as given by (\ref{o-28}). To sum up them we write the second equality in 
(\ref{o-111}) in the form 
$\sum_{l=1}^{\No}\t{C}'_{j,l}\t{F}(x,\ka_{l},\mu)=-\t{C}'_{j,\No+1}\t{F}(x,\ka_{\No+1},\mu)$, so that 
$\sum_{l=1}^{\No}\t{C}'_{j,l}\t{F}(x,\ka_{l},\ka_{\No+1})=-\t{C}'_{j,\No+1}\t{F}(x)$ thanks to (\ref{o-112}). Then 
 \begin{equation}
\sum_{l=1}^{\No}\t{C}'_{jl}\dfrac{\t{r}_{l}}{r_{l}}F(x,\ka_l,\mu)=-\t{r}_{\No+1}\t{C}'_{j,\No+1}\dfrac{\t
{ F} (x , \ka_ { \No+1},\mu)}
{\t{F}(x)-r_{\No+1}},\label{o-30}
 \end{equation} 
 where $j=1,\ldots,N_a+1$.  Thus in order to get zero in the r.h.s.\ for $j=1,\ldots,N_a$ we have first to reduce
matrix $\t{C}'$ by means of  (\ref{gr}) to the form where the last column has all zeros with exception to the last element. 
Then matrix $C'$ is given by omitting the last column and last row of the matrix $\t{C}'$ and multiplication by the 
diagonal matrix $\diag{r_n/\t{r}_{n}}$ from the right.

Derivation of the $B^{-1}$-transformation is performed in the same way. We assume that $\t{F}(x,\la,\mu)$ with parameters 
$\ka_{1},\ldots,\ka_{\No+1}$ ($\t{N}_{a}=N_a$, $\t{N}_{b}=N_b+1$) and matrix $\t{C}$ (or $\t{C}$) is given and we use 
(\ref{o-66}) to define $(N_a,N_b)$-soliton function $F(x,\la,\mu)$ with parameters $\ka_{1},\ldots,\ka_{\No}$. Setting in 
(\ref{o-66}) $\la=\ka_{\No+1}$ we get due to (\ref{o-112})
\begin{equation}
\dfrac{F(x,\ka_{\No+1},\mu)}{R(\mu)}=\dfrac{\t{F}(x,\ka_{\No+1},\mu)\t{r}_{\No+1}}{\t{R}(\mu)\t{F}(x)},
\label{o-68}
\end{equation} 
so that under this substitution (\ref{o-66}) reads as
\begin{equation}
\dfrac{F(x,\la,\mu)}{R(\mu)}=\dfrac{1}{\t{R}(\mu)}\biggl[\t{F}(x,\la,\mu)-\dfrac{\t{F}(x,\la,\ka_{\No+1})\t{F}(x,\ka_{\No+1},
\mu)}{\t{F}(x)}\biggr].\label{o-69}
\end{equation} 
We see that the pole at $\mu=\ka_{\No+1}$ in the r.h.s.\ (see (\ref{tA})) is compensated thanks to (\ref{o-112}) by a zero 
of numerator at this point, that proves that $\ka_{\No+1}$ is excluded from parameters of the function $F$. Now for
$l,m=1,\ldots,\No$
\begin{align}
&\dfrac{F(x,\ka_{l},\mu)}{R(\mu)}=\dfrac{1}{\t{R}(\mu)}\biggl[\t{F}(x,\ka_{l},\mu)-\dfrac{\t{F}(x,\ka_l,\ka_{\No+1})}{\t{F}
(x)}\t{F}(x, \ka_ {\No+1},\mu)\biggr],\label{o-72}\\
&\dfrac{F(x,\la,\ka_{m})}{r_{m}}=\dfrac{1}{\t{r}_{m}}\biggl[\t{F}(x,\la,\ka_{m})-\t{F}(x,\la,\ka_{\No+1})\dfrac{\t{F}(x,
\ka_{\No+1},\ka_m)}{\t{F}(x)}\biggr],\label{o-73}
\end{align}
where (\ref{Rr}) and its tilde version were used. Thus, thanks to the tilde version of (\ref{f41}) we have that
\begin{equation}
\sum_{l=1}^{\No}\t{C}'_{jl}F(x,\ka_{l},\mu)=0,\quad j=1,\ldots,N_a,\label{o-74} 
\end{equation} 
that means that 
\begin{equation}
C'=\Vert\t{C}'_{jl}\Vert_{\substack{j=1,\ldots,N_a\\ l=1,\ldots,\No}},\label{o-75} 
\end{equation} 
i.e., transformed matrix $C'$ equals matrix $\t{C}'$ with the last column omitted. Next, by (\ref{o-73}) and tilde 
version of (\ref{f42})
\begin{equation}
\sum_{m=1}^{\No}F(x,\la,\ka_{m})\dfrac{\t{r}_{m}}{r_{m}}\t{C}_{mj}=-\dfrac{\t{r}_{\No+1}\t{F}(x,\la,\ka_{\No+1})}{\t{F}(x)}\t
{C}_{\No+1,j},\label{o-76} 
\end{equation} 
where $j=1,\ldots,N_b+1$. By means of (\ref{gr}) we reduce matrix $\t{C}$ to the form where the last row
equals ($0,\ldots,0,1$). Then the transformed matrix $C$ obeying (\ref{f42}) is given by means of relation
\begin{equation}
rC=\Vert(\t{r}\t{C})_{lj}\Vert_{\substack{j=1,\ldots,N_b\\ l=1,\ldots,\No}},\label{o-77}
\end{equation} 
where we used notation (\ref{r}). Thus the transformed matrix $rC$ is given by the reduced as above matrix 
$\t{r}\t{C}$ with the last column and last row being omitted. It is easy to see that matrices $C$ and $C'$ obey (\ref{d12}) 
because matrices $\t{C}$ and $\t{C'}$ obey the tilde version of this equality.

\section{Discussion} \label{discussion}
In  our work \cite{BPP2015}, where the famous KPII equation was used as an example, we demonstrated that the Cauchy--Jost 
function is the very natural tool for the Inverse problem. Here we developed this approach by including the case of soliton 
solutions. The Cauchy--Green function was defined here, not in terms of the Jost solutions (see (\ref{new1})) but 
directly by its analyticity properties with respect to the spectral parameters, see Definition~\ref{def1}. We proved that in 
the KPII case relation  (\ref{new1}) results from this definition. We derived and presented properties of the Cauchy--Jost 
function in detail and gave linear equations defining Darboux transformations. We proved that these transformations are 
naturally decomposed in three classes: transformations of $A$, $B$ and 0 types, 
given by (\ref{o-549}),  (\ref{o-66}) and (\ref{01}) with auxiliary conditions (\ref{o-551}),  (\ref{o45}) and (\ref{02}) 
correspondingly. This classification is generic and relations  (\ref{o-549}),  (\ref{o-66}) and (\ref{01}) define both, 
direct and inverse transformations. 

Well known problem of the description of soliton solutions of the KPII equation is the problem of their regularity on 
the $(x_1,x_2)$-plane for any $x_3$, that thanks to (\ref{u}) is equivalent to the absence of zeros of the 
$\tau$-function. Without loss of generality we can assume that
\begin{equation}
 \ka_{1}<\ka_{2}<\ldots<\ka_{\No},\label{kappa}
\end{equation}
that gives by (\ref{Vo}) that all maximal minors $\Vo({n_i})$, $n_1<n_2<\ldots<n_{N_b}$, in (\ref{tauf1}) of the Vandermond 
matrix $\Vo$ are positive. Then for the positivity of $\tau$-function in (\ref{tauf1}) it is enough to impose condition that 
the same is property of the maximal minors of the matrix $C$:
\begin{equation}
C(\{n_i\})\geq0,\quad n_1<n_2<\ldots<n_{N_b},
\end{equation}
i.e., that matrix $C$ is totally nonnegative (TNN). Necessity of this condition was proved in \cite{KW}. In terms of the 
Cauchy--Jost function regularity of the potential $u(x)$ is equivalent to regularity of this function itself thanks to 
relations (\ref{new13}). It is easy to see that the Darboux transformations do not guarantee that function $\t{F}(x,\la\mu)$ 
given by any of direct transformations is regular, if the regularity of the initial function ${F}(x,\la\mu)$ is assumed. 
While for the inverse $A$-transformation we get that TNN property of matrix $C$ is preserved under reduction 
procedure given after (\ref{o-29}). The same for the inverse $B$-transformation for matrix $C'$ as follows from 
(\ref{o-75}). This proves that an arbitrary regular $(N_a,N_b)$-soliton solution can be reached from the zero one by 
successive direct Darboux transformations that preserves regularity property on every step. 

On the other side, the results of direct transformations, relations (\ref{o-1}) and (\ref{o-11}) together with properties of 
the Cauchy--Jost function given in Sec.\ 3, can be used as a tool to control absence of singularities in the transformed 
objects. Let us consider example of $A$-transformation, i.e., equality (\ref{o-1}).  Assuming regularity of the original 
potential $u(x)$ and thus the original Cauchy--Jost function $F(x,\la,\mu)$, we see that the regularity of the transformed 
function $\t{F}(x,\la,\mu)$ is equivalent to the absence of zeros in the denominator in the r.h.s. Thus by (\ref{new13}), 
(\ref{new151}) and taking notations (\ref{Kn}),  (\ref{Rla}) and (\ref{new8}) into account we have that
\begin{align}
 &\sum_{l=1}^{\No}\dfrac{a_l}{r_{l}}F(x,\ka_{l},\ka_{\No+1})-a_{\No+1}=\nonumber\\
&=\dfrac{1}{\tau(x)}\sum_{1\leq{n}_1<\ldots<n_{N_b}\leq\No}C(\{n_{i}\})\Vo(\{n_{i}\})\prod_{j=1}^{N_{b}}e^{\ell_{n_{j}}x}_
{}(\ka_{\No+1}-\ka_{n_j})\times\nonumber\\
&\times\sum_{k=1}^{N_b}\dfrac{a_{n_k}e^{(\ell_{\No+1}-\ell_{n_k})x}_{}}{(\ka_{\No+1}-\ka_{
n_k } )\displaystyle\prod_{i=1,\,i\neq{k}}^{N_b}(\ka_{n_k}-\ka_{n_i})}-a_{\No+1}.\label{tnn1}
\end{align}
By assumption, $\tau(x)$ is positive, that means that the whole factor in the second line is positive as well (we use 
Condition~\ref{cC} here). Now, thanks to (\ref{kappa}) sign of the product in the last line equals to $(-1)^{N_b+k}$ and in 
generic situation one cannot make the sum  in the third line to be sign determined by choice of parameters $a_k$. But 
let us assume that,say, only the last parameter in the set $\{a_{1},\ldots,a_{\No}\}$ is different from zero. 
$a_n=a\delta_{n,\No}$. Then $n_{N_b}=\No$ due to condition on the indexes of summation in (\ref{tnn1}), so that 
\begin{align}
 &\sum_{l=1}^{\No}\dfrac{a_l}{r_{l}}F(x,\ka_{l},\ka_{\No+1})-a_{\No+1}=\nonumber\\
&=\dfrac{ae^{\ell_{\No+1}x}_{}}{\tau(x)}\sum_{1\leq{n}_1<\ldots<n_{N_b-1}\leq\No-1}C(\{n_{i},\No\})\Vo(\{n_{i},\No\})\prod_{
j=1}^{N_{b}-1}e^{\ell_{n_{j}}x}_
{}\dfrac{\ka_{\No+1}-\ka_{n_j}}{\ka_{\No}-\ka_{n_i}}-\nonumber\\
&-a_{\No+1},\label{tnn2}
\end{align}
so that choice $\sgn{a_{\No+1}}=-\sgn{a}$ guaranties that all terms in the r.h.s.\ have the same sign, and thus the 
resulting soliton solution is regular. Correspondingly, by \cite{KW} this means that matrix $\t{C}$ given in (\ref{o-615}) is 
totally nonnegative. The problem of how generic are TNN matrices constructed in this way deserves further investigation. 

\appendix
\section{Proof of relations  (\ref{new6}) and (\ref{tautau})}\label{app1} 
By definition $\No\times{N_b}$ real matrix $C$ obeys Condition~\ref{cC}, thus it can be written in the form 
\begin{equation}
C=p\left(\begin{array}{c}I_{N_b}\\d\end{array}\right)c,\label{appen1}  
\end{equation} 
where $p$ is a permutation matrix, $I_{N_b}$ is the unity $N_b\times{N_{b}}$-matrix, $d$ is an 
$N_a\times{N_{b}}$-matrix and $N_b\times{N_{b}}$-matrix $c$ is nondegenerate. Then relation 
(\ref{d12}) gives that there exists such $N_a\times{N_{a}}$-matrix $c'$ that
\begin{equation}
C'r=c'\bigl(-d,I_{N_a}\bigr)p^{-1}.\label{appen2} 
\end{equation} 
Thanks to this any maximal minor of the matrix $C'r$ is proportional to $\det{c'}$, thus by condition matrix $c'$ is 
nonsingular. Then $C^{\text{T}}C=c^{\text{T}}(I_{N_b}+d^{\text{T}}d)c$ and 
$C'r^{2}{C'}^{\text{T}}=c'(I_{N_a}+dd^{\text{T}}){c'}^{\text{T}}$, that proves that both these 
products are positive (and thus invertible) matrices.  Let us introduce  $\No\times\No$ matrices 
\begin{equation}
\pi=C({C}^{\text{T}}C)^{-1}C^{\text{T}},\qquad\pi^{\,\prime}=r{C'}^{\text{T}}(C'r^{2}{C'}^{\text{T}})^{-1}C'r. \label{d18}
\end{equation}
Thanks to this definition and (\ref{d12}) these matrices are orthogonal projectors:
\begin{equation}
\pi^{2}=\pi,\qquad {\pi'}^{2}=\pi',\qquad \pi\pi'=0=\pi'\pi, \label{d22}
\end{equation}
and because of the above definitions we have for them
\begin{align}
p^{-1}\pi{p}&=\left(\begin{array}{cc}
(I_{N_b}+d^{\text{T}}d)^{-1},&(I_{N_b}+d^{\text{T}}d)^{-1}d^{\text{T}}\\
d(I_{N_b}+d^{\text{T}}d)^{-1},&d(I_{N_b}+d^{\text{T}}d)^{-1}d^{\text{T}}\end{array}\right),\label{appen3}\\
p^{-1}\pi'{p}&=\left(\begin{array}{cc}
d^{\text{T}}(I_{N_a}+dd^{\text{T}})^{-1}d,&-d^{\text{T}}(I_{N_a}+dd^{\text{T}})^{-1}\\
-(I_{N_a}+dd^{\text{T}})^{-1}d,&(I_{N_a}+dd^{\text{T}})^{-1}\end{array}\right).\label{appen4} 
\end{align}
Using now obvious relations of the kind 
\begin{align*}
&(I_{N_b}+d^{\text{T}}d)^{-1}+d^{\text{T}}(I_{N_a}+dd^{\text{T}})^{-1}d=\\
&\qquad=(I_{N_b}+d^{\text{T}}d)^{-1}[I_{N_b}+(I_{N_b}+d^{\text{T}}d)d^{\text{T}}(I_{
N_a}+dd^{\text{T}})^{-1}d]=\\
&\qquad=(I_{N_b}+d^{\text{T}}d)^{-1}[I_{N_b}+d^{\text{T}}(I_{N_a}+dd^{\text{T}})(I_{N_a}+dd^{\text{T}})^{-1}d]=I_{N_b},
\end{align*}
we prove by (\ref{appen3}) and (\ref{appen4}) that projectors $\pi$ and $\pi'$ are complementary in the sense that 
\begin{equation}
\pi+\pi'=E_{\No},\label{d23} 
\end{equation} 
where $E_{\No}$ is $\No\times\No$ unity matrix. This gives equality (\ref{new6}), where 
\begin{equation*}
w_{b}=({C}^{\text{T}}C)^{-1}C^{\text{T}}w\text{ and }w_{a}=(C'r^{2}{C'}^{\text{T}})^{-1}C'rw. 
\end{equation*}

In order to prove Eq.\ (\ref{tautau}) we write the Vandermonde matrix $V$ (see (\ref{V})) in the form
$\Vo=v(u,I_{N_a})p^{-1}$, where $p$ is the same permutation matrix as in (\ref{appen1}) , $v$ and $u$ are some 
$N_a\times{N_a}$ and $N_a\times{N_b}$ matrices, correspondingly, defined by this equality. Then by (\ref{d12'}) 
\begin{equation*}
r^{-1}\Vo'=p\left(\begin{array}{c}I_{N_b}\\-u\end{array}\right)v',\end{equation*}
where $v'$ is a $N_b\times{N_b}$ matrix. By definition both matrices $v$ and $v'$ are nonsingular.
Let us introduce two diagonal matrices $e_{N_a}^{\ell{x}}$ and $e_{N_b}^{\ell{x}}$ defined by the 
equality (cf.\ (\ref{eK}))
\begin{equation*}
p^{-1}e_{}^{\ell{x}}p=\left(\begin{array}{cc}
e_{N_b}^{\ell{x}}& 0\\
0& e_{N_a}^{\ell{x}}\end{array}\right).
\end{equation*}
Then using (\ref{appen1}) we can write the first equality in (\ref{tau}) in the form
$\tau(x)=\det\bigl(v(ue_{N_b}^{\ell{x}}+e_{N_a}^{\ell{x}}d)c\bigr)$. Finally we get
\begin{align*}
&\tau(x)=\det{v}\det{c}\det{e_{N_b}^{\ell{x}}}\det\bigl(u+e_{N_a}^{\ell{x}}
de_{N_b}^{-\ell{x}}\bigr),\\
&\tau'(x)=(-1)^{N_b}\det{v'}\det{c'}\det{e_{N_a}^{-\ell{x}}}\det\bigl(u+e_{N_a}^{\ell{x}}
de_{N_b}^{-\ell{x}}\bigr),
\end{align*}
where the second equality is derived in analogy. These relations prove (\ref{tautau}).

\end{document}